\documentclass[fleqn,usenatbib,useAMS]{mnras}

\usepackage{graphicx}	
\usepackage{amsmath}	
\usepackage{amssymb}	
\usepackage{multicol}   
\usepackage{bm}		    
\usepackage{pdflscape}	
\usepackage{xcolor}	    
\usepackage{subcaption}
\usepackage{caption}
\usepackage[T1]{fontenc}
\usepackage{ae,aecompl}

\usepackage{newtxtext,newtxmath}

\title[Novel Constraints on Companions to the Helix]{Novel Constraints on Companions to the Helix Nebula Central Star}

\author[Iskandarli et al.]{Leyla Iskandarli,$^1$\thanks{E-mail:leyla.iskandarli.21@ucl.ac.uk}
Jay Farihi,$^1$
Joshua D.~Lothringer,$^2$
Steven G.~Parsons,$^3$
Orsola De Marco,$^4$
\newauthor
and
Thomas Rauch$^5$
\\
$^1$Department of Physics and Astronomy, University College London, London WC1E 6BT, UK\\
$^2$Space Telescope Science Institute, Baltimore, MD 21218, USA\\
$^3$Department of Physics and Astronomy, University of Sheffield, Sheffield S3 7RH, UK\\
$^4$School of Mathematical and Physical Sciences, Macquarie University, Sydney, NSW 2109, Australia\\
$^5$Institut f\"ur Astronomie und Astrophysik, Kepler Center for Astro and Particle Physics, Eberhard Karls Universit\"at, 72076 T\"ubingen, Germany
}

\date{Accepted XXX. Received YYY; in original form ZZZ}

\pubyear{2024}

\begin{document}
\maketitle

\begin{abstract}
The Helix is a visually striking and the nearest planetary nebula, yet any companions responsible for its asymmetric morphology have yet to be identified.  In 2020, low-amplitude photometric variations with a periodicity of 2.8\,d were reported based on Cycle 1 {\em TESS} observations. In this work, with the inclusion of two additional sectors, these periodic light curves are compared with {\sc lcurve} simulations of irradiated companions in such an orbit.  Based on the light curve modelling, there are two representative solutions: i) a Jupiter-sized body with 0.102\,R$_\odot$ and an arbitrarily small orbital inclination $i=1\degr$, and ii) a 0.021\,R$_\odot$ exoplanet with $i\approx25\degr$, essentially aligned with the Helix Nebular inclination.  Irradiated substellar companion models with equilibrium temperature 4970\,K are constructed and compared with existing optical spectra and infrared photometry, where Jupiter-sized bodies can be ruled out, but companions modestly larger than Neptune are still allowed.  Additionally, any spatially-unresolved companions are constrained based on the multi-wavelength, photometric spectral energy distribution of the central star.  No ultracool dwarf companion earlier than around L5 is permitted within roughly 1200\,au, leaving only faint white dwarfs and cold brown dwarfs as possible surviving architects of the nebular asymmetries.  While a planetary survivor is a tantalizing possibility, it cannot be ruled out that the light curve modulation is stellar in nature, where any substellar companion requires confirmation and may be possible with {\em JWST} observations.
\end{abstract}




\section{Introduction}

Planetary nebulae represent a transient phase in the life cycle of intermediate-mass stars, occurring after they exhaust their helium fuel and evolve into red giants for the second time. During this phase, the star sheds its outer layers through stellar winds, forming a nebula illuminated by the hot, exposed core, which subsequently becomes a white dwarf. The Helix Nebula (NGC\,7293), one of the closest and most studied planetary nebulae, provides a unique opportunity to explore the characteristics of its central star and potential companions \citep{Balick2002}.

Over the past few decades it has become clear that binary systems can lead to the formation of asymmetries in planetary nebulae, and are likely the main culprits in shaping all but the spherical or mildly elliptical nebular morphologies \citep{Soker1997,DeMarco2009,Jones2017}. While the influence of binary systems on the evolution of planetary nebulae is well documented, the fraction of those that contain a sufficiently close binary to have interacted in a common envelope, and the nature of the companions within these systems, remain comparatively less constrained. Currently, there are about 60 known binary systems in planetary nebulae \citep{Boffin2019}, where roughly 20\,per cent of these are consistent with post-common envelope systems \citep{Bond2000, Miszalski2009}. 

In planetary nebulae, the majority of close binary systems are detected by observing light curve fluctuations, highlighting the significance of this method for identifying stellar companions \citep{Jacoby2021}. These fluctuations, characterised by dimming and brightening patterns, indicate binary systems via eclipses, irradiation of the companion by the central star, or ellipsoidal variation as the central star tidally distorts the secondary. Photometric variability analysis, best if accompanied by radial velocity curves, can then retrieve a number of orbital and stellar characteristics \citep[e.g.][]{Hillwig2016}.

The possibility of planetary mass companions that may have influenced the nebular morphology has been previously studied \citep[e.g.][]{Soker1998a,DeMarco2011}. Though technically feasible, it is challenging for companions with masses as low as several M$_{\rm Jup}$ to have survived a strong interaction with the asymptotic giant branch (AGB) progenitor of the central star \citep{Passy2012}. There are a handful of known brown dwarf companions to white dwarfs, with orbital periods of hours that signify common envelope evolution \citep[eight systems are listed in a compendium;][]{Zorotovic2022}.  However, the masses of these substellar survivors are estimated to be in the range $50-70$\,M$_{\rm Jup}$, where less massive companions are thought to be lost in the engulfment process \citep{Walters2023}.  Any indication of lower-mass survivors of a common envelope phase -- as are close companions to planetary nebula central stars -- would place strong constraints on the efficiency of envelope ejection and exoplanet fates during post-main sequence evolution \citep[e.g.][]{Mustill2012,Nordhaus2013}.

In 2020, using sensitive space-based observations from the {\em Transiting Exoplanet Survey Satellite} \citep[{\em TESS};][]{ricker2015} during Cycle 1, seven out of eight planetary nebulae central stars were found to have periodically variable light curves \citep{Aller2020}. These eight systems represented the only such targets with 2\,min cadence observations at that time, and thus these results suggest a high candidate binary fraction among central stars; cf.\ a binary fraction of 21~per cent found using {\em K2} \citep{Jacoby2021}, but strong hints of a higher fraction using observations from the original {\em Kepler} mission \citep{DeMarco2015}.  The study by \citet{Aller2020} reported periodic signals ranging from 1.7 to 6.8\,d, where all seven variable light curves display morphologies consistent with the effects of an irradiated companion, with two showing further modulation as expected from tidal distortion.  

The  central star of the Helix planetary nebula was found to exhibit a 2.77\,d period based on Sector 2 {\em TESS} observations, but the light curve modeling was not successful within the range of available dwarf stellar templates using {\sc phoebe 2.0} \citep{Prsa2016}.  The companions tested spanned the range from the latest main-sequence B-type stars (2.5\,M$_\odot$) through approximately an M5-type dwarf (0.16\,M$_\odot$), and it was speculated that lower-mass stellar or substellar companions remained plausible \citep{Aller2020}.

Building on previous work, this study takes a comprehensive approach to constrain companions to the Helix Nebula central star, including an independent light curve modeling effort and consideration of the multi-wavelength spectral energy distribution (SED).  Using two new sectors of {\em TESS} data, this paper provides an improved ephemeris, as well as companion parameters for two sets of possible solutions based on simulated light curves.  A set of irradiated substellar atmosphere models are generated for comparison with current and future data, and broad constraints are placed on all possible spatially-unresolved companions that might be hidden architects of the nebular asymmetries.

Section \ref{sec:2} details the observational data and subsequent analyses, including modelling of the light curve and SED. Section \ref{sec:3} presents the resulting constraints on potential companions, as well as models for an irradiated substellar atmosphere that could be responsible for the observed light curve, and compares these predictions to existing data.  Finally, Section \ref{sec:4} summarizes the findings and suggests future work to constrain companions to the Helix.

\section{Data and analysis}
\label{sec:2}

\subsection{Time-series and multi-wavelength photometry}

Time-series data for the Helix central star were collected by {\em TESS}, with a 2\,min cadence over Sectors 2, 28, and 42. These observations were processed by the Science Processing Operations Center, and the {\sc pdcsap} data were obtained from the Mikulski Archive for Space Telescopes. The light curves were processed by applying a $5\upsigma$ clipping to exclude outliers, which removed approximately 0.01\,per cent of points, followed by the removal of NaN entries. The remaining data were then analysed using Lomb-Scargle and Fourier periodograms.

The data utilised for constructing the photometric SED were obtained from multiple sources. Ultraviolet through infrared photometry were taken from {\em GALEX}, Pan-STARRS, and 2MASS \citep{martin2005,skrutskie2006,tonry2012}, and via pointed observations using {\em Spitzer} IRAC \citep{Su2007}. In the case of {\em GALEX} photometry, the source catalogue was eschewed in favour of dedicated measurements made for central stars of planetary nebulae \citep{Gomez-Munoz2023}.

\subsection{Central star parameters}

\begin{table}
\begin{center}
\caption{Adopted parameters for the Helix central star.}
\label{tab:Tab1}
\begin{tabular}{lr}
		
\hline

Parameter					&Value\\

\hline

Spectral Type				&DAO\\
$V$ (mag)					&$13.524\pm0.002$\\
Distance (pc)				&$199.5\pm1.7$\\
$T_{\rm eff}$ (K)			&$120\,000\pm6000$\\
Mass (M$_{\odot}$)			&$0.678\pm0.025$\\
Radius (R$_{\odot}$)		&$0.025\pm0.001$\\

\hline

\end{tabular}
\end{center}
\end{table}

The analysis of the Helix central star is based on the following adopted parameters. The central star is categorized as a DAO-type white dwarf with a hydrogen-dominated atmosphere \citep{Napiwotzki1999, Gianninas2011}, where an effective temperature of $T_{\rm eff} = 120\,000\,\pm\,6000$\,K has been determined via detailed modeling of myriad ultraviolet and optical spectral lines, including 21 metal species \citep{Traulsen2005, Ziegler2013}.  The {\em Gaia} EDR3 distance to the Helix is 200\,pc \citep{Gaia2021}, and the nebula itself has an inclination of $28\degr\pm10\degr$, as determined by fitting an ellipse to the H$\upalpha$ surface brightness contours out to 0.6\,pc from the central star \citep{Henry1999}.

While there are several studies of the Helix that determine fundamental stellar parameters, all of these are pre-{\em Gaia} \citep[e.g.][]{Benedict2009}, and none of those utilize the available ultraviolet spectroscopy.  Adopting $T_{\rm eff} = 120\,000$\,K and the {\em Gaia} parallax $\varpi=5.01\pm0.04$, the apparent $BVRI$ magnitudes of the Helix central star \citep{Landolt2007} imply absolute magnitudes that correspond to a surface gravity with an average and standard deviation $\log [g({\rm cm\,s^{-2}})]=7.466\,\pm0.003$ \citep{Bedard2020}.  This $T_{\rm eff}$ and $\log g$ corresponds to a mass and radius of $M=0.68$\,M$_\odot$ and $R=0.025$\,R$_\odot$\footnote{\url{https://www.astro.umontreal.ca/~bergeron/CoolingModels/}}.  Table~\ref{tab:Tab1} summarizes the adopted parameters for the central star, where the errors in mass and radius are calculated by propagating the uncertainties in $T_{\rm eff}$ and the {\em Gaia} distance into the evolutionary models for the corresponding absolute magnitudes.

\begin{figure*}
    \includegraphics[width=\columnwidth]{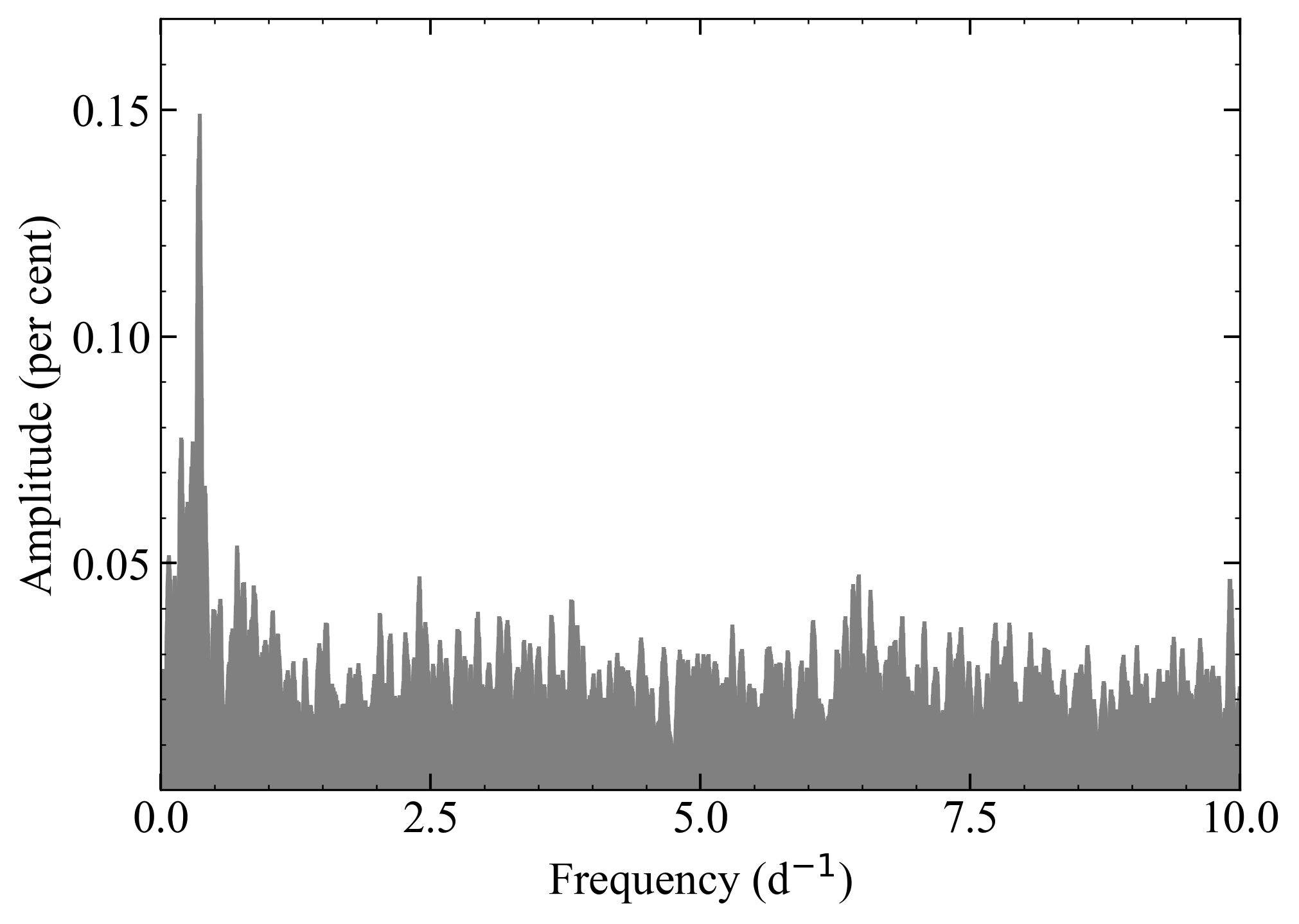}  
    \includegraphics[width=\columnwidth]{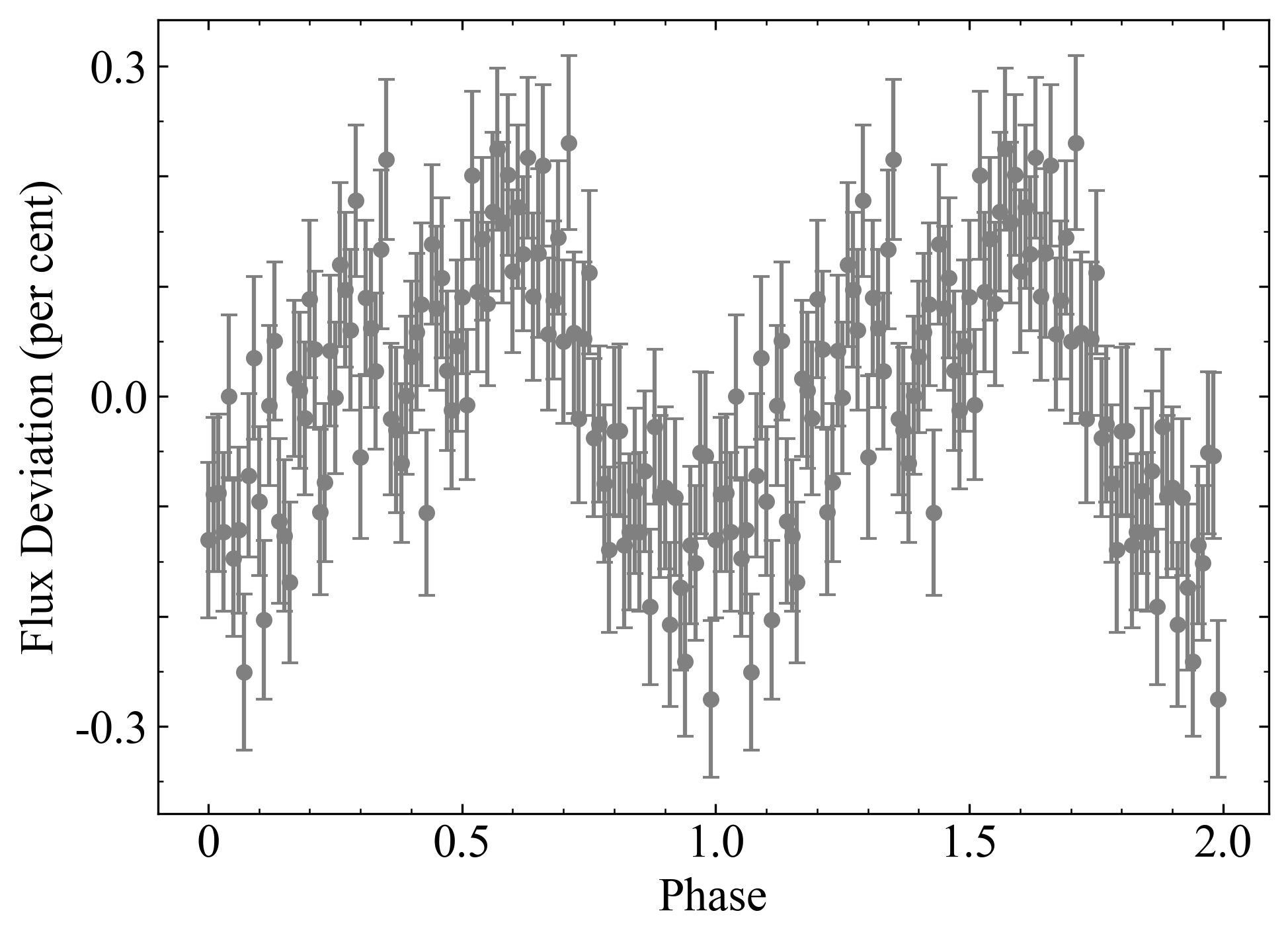} 
    \caption{On the left is a Fourier periodogram for the combined Sectors 2, 28, and 42 of {\em TESS} observations, where the strongest peak occurs at 0.3582\,d$^{-1}$, with amplitude 0.15\,per cent.  On the right is the resulting light curve for these data, phase-folded on the peak periodogram frequency into 100 evenly-spaced phase bins, and shown as the per cent deviation from the mean.}
    \label{fig:combined_figure}
\end{figure*}

\subsection{Light curve analysis}

To improve the modelling precision of the Helix central star light curve, the {\em TESS} data were analyzed using {\sc period04} \citep{lenz2005}.  The most significant peak in the periodogram occurs at 2.79\,d, and is within $1\upsigma$ of the previously published frequency based on the first sector observed \citep{Aller2020}.  Errors in frequency, phase, and amplitude were established using Monte Carlo simulations, where the revised ephemeris is now two orders of magnitude more precise in frequency. The new ephemeris, including three {\em TESS} sectors is:\\

\noindent
BJD$_{\rm TDB} = 2458355.33(3) + 2.7915(2)\,E$\\

\noindent
The ephemeris is given such that phase 0 corresponds to the light curve minimum, where $T_0$ is selected to be the minimum nearest the start of Sector 2 {\em TESS} observations.  Both the periodogram and phase-folded light curve are shown in Figure~\ref{fig:combined_figure}.  

As an additional test for a consistent and stable signal, the three {\em TESS} sectors were analyzed individually using {\sc period04}.  Sectors 2 and 28 yield peak frequencies with errors of $0.3627\pm0.0020$ and $0.3507\pm0.0025$\,d$^{-1}$, respectively, which nominally differ by $3\upsigma$.  Sector 42 exhibits a periodogram peak at 0.348\,d$^{-1}$, but with an error that is at least 20\,per cent, and thus consistent with the other two sectors.  On further investigation, the total observational baseline of the Helix is 26.0, 19.3, and 13.7\,d (9.4, 7.0, and 4.9 cycles) in Sectors 2, 28, and 42, respectively.  Based on this, the mild disagreement is likely a result of stunted coverage, the low-amplitude signal, where {\sc pdcsap} de-trending may also affect individual periodogram peaks at low frequencies.

\subsection{Spectral energy distribution}

\begin{figure}
    \includegraphics[width=\columnwidth]{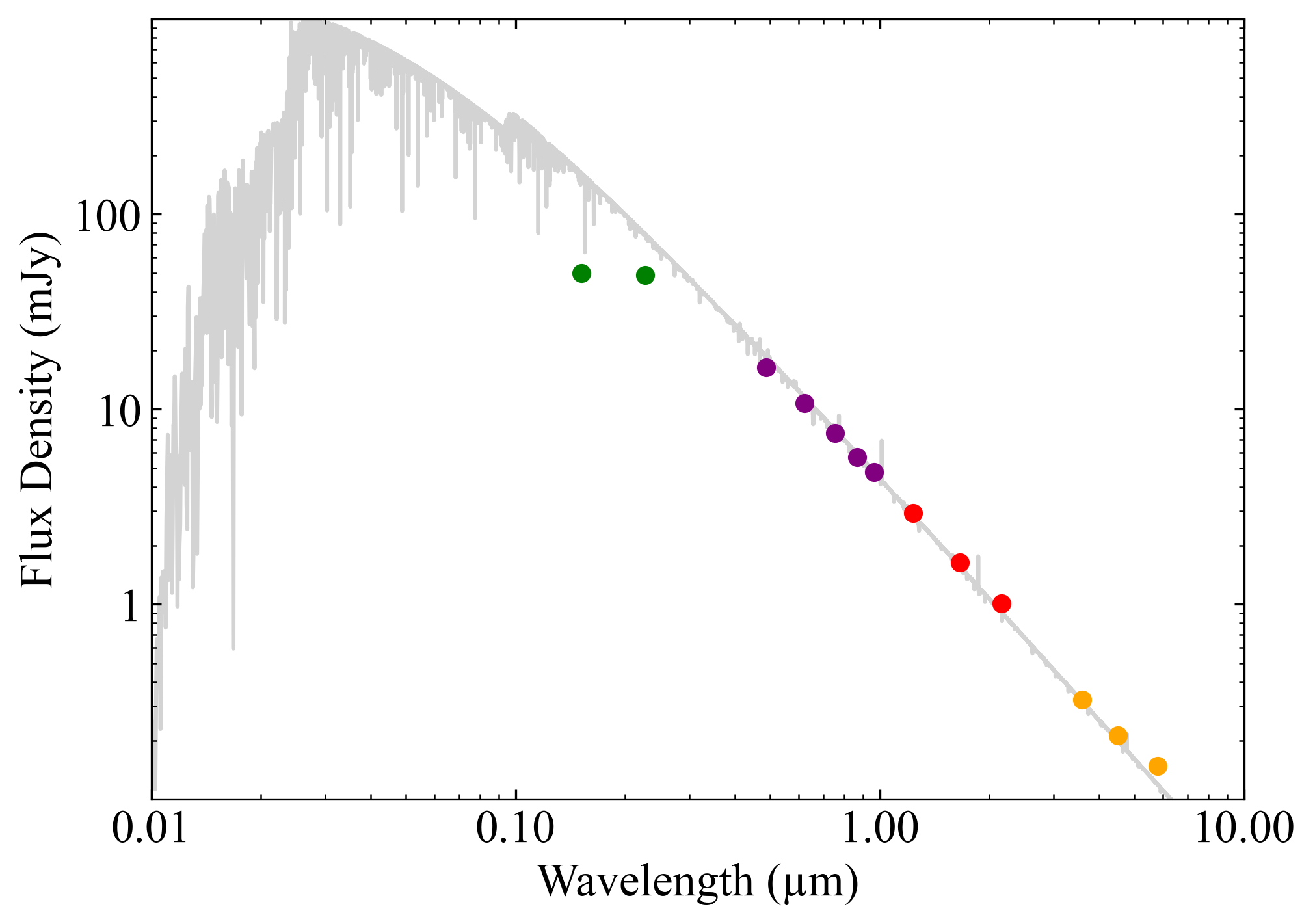}
    \caption{The SED of the Helix Nebula white dwarf.  Plotted in grey is the model stellar atmosphere, based on the parameters in Table~1, and coloured circles are observed photometry: {\em GALEX} (green), Pan-STARRS (purple), 2MASS (red), and {\em Spitzer} IRAC (gold). The IRAC $7.9\,\upmu$m photometry is excluded from analysis owing to the emerging excess from cold dust at this wavelength \citep{Su2007}.  Photometric error bars are present but smaller than the plotted data points.}
    \label{fig:SED}
\end{figure}

The photometric SED of the central star is well constrained in the infrared and sensitive to unevolved companions such as low-mass stars and brown dwarfs, which tend to be bright in the $4-5\,\upmu$m range \citep{burrows2003,marley2021}.  These data are plotted with a model of the white dwarf photosphere in Figure~\ref{fig:SED} out to $5.7\,\upmu$m; beyond this wavelength, the Helix is known to exhibit an infrared excess consistent with cool dust \citep{Su2007}.  There is no obvious photometric excess at these wavelengths, indicating any potential companion must remain undetected in these data.  Generally speaking, a lack of photometric excess constrains both the effective temperature and radius of potential companions.  For this study, a detectability threshold of 10\,per cent is adopted, which is essentially at or above $5\upsigma$ for a typical calibration-limited observation with {\em Spitzer} IRAC in all cryogenic bandpasses \citep{Reach2005}. 

First, any companion in a 2.79\,d orbit will be prone to extreme irradiation by the central star, and should have an equilibrium temperature of 4970\,K (assuming no internal heat sources).  This temperature is roughly equivalent to that of a K3-type main-sequence star \citep{Pecaut2013}, which would outshine the white dwarf in the infrared (e.g.\ by nearly four magnitudes in the $K$ band). Thus, to remain undetectable, any such heated companion must have a significantly smaller radius than a main-sequence star at this temperature.  Consequently, based purely on photometry, the upper radius limit for any irradiated companion in a 2.79\,d orbit is 0.102\,R$_\odot$.  A re-scaled K3 dwarf star with this smaller radius can be added to the photometric SED of the white dwarf, and all infrared fluxes remain below a 10\,per cent excess threshold.

Second, the published infrared flux measurements for the Helix central star \citep{Su2007} are compared against predictions for the white dwarf atmospheric models plus a range of {\em non-irradiated} ultracool dwarf companions.  Such potential companions could remain spatially unresolved and be undetected to date by any means, where {\em Spitzer} IRAC provides the strongest constraints for objects within its photometric aperture.  While the standard photometric aperture for IRAC is 10 native pixels or 12\,arcsec, a more conservative 6\,arcsec is adopted for potentially unresolved companions, which corresponds to a projected distance of 1200\,au at the {\em Gaia} distance to the Helix.

\begin{figure}
\centering
    \includegraphics[width=0.95\columnwidth]{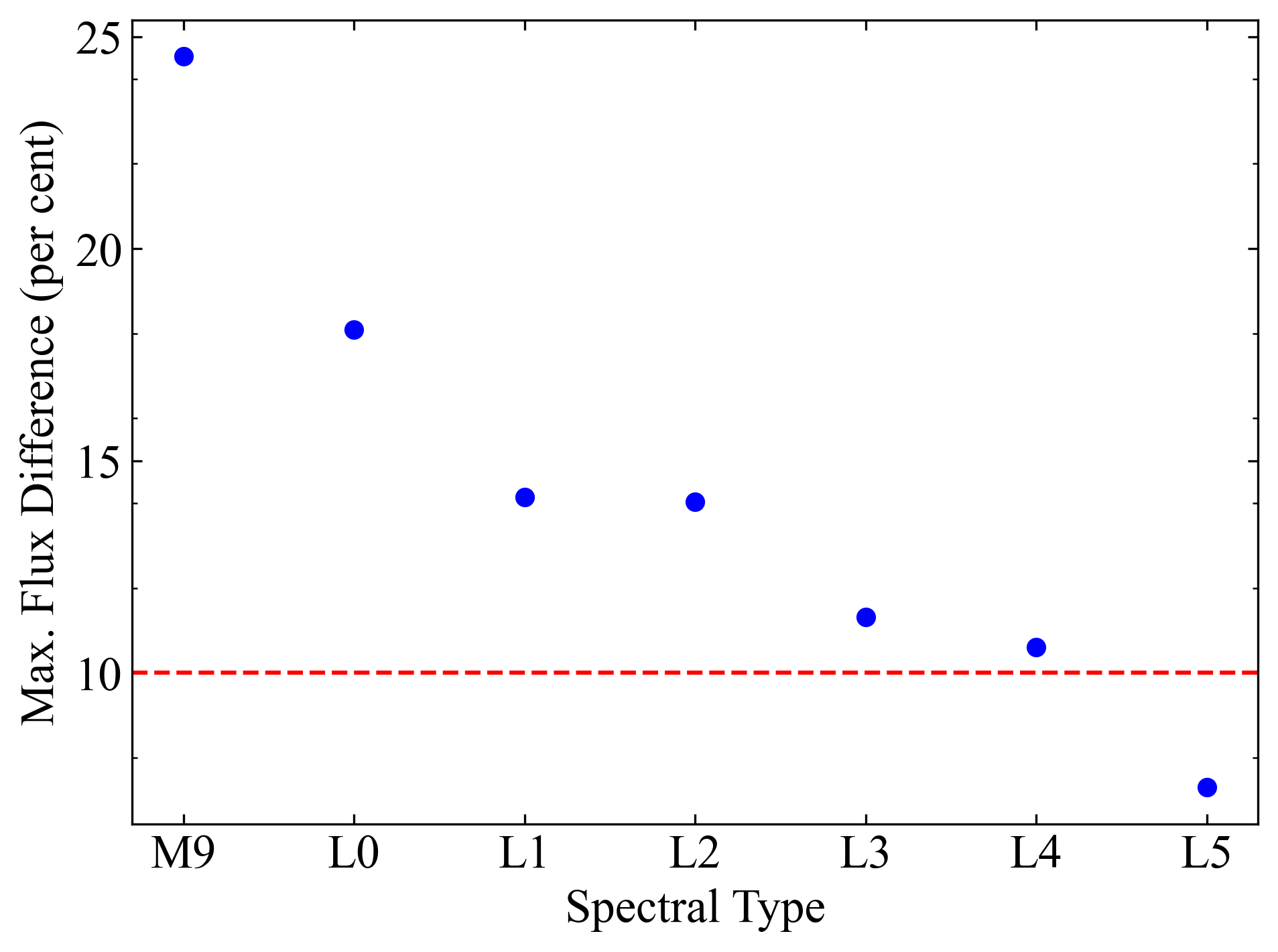}
    \caption{Maximum flux difference between the measured photometry and with the addition of an ultracool dwarf companion to the white dwarf model photopshere, as a function of spectral type.  These two fluxes are compared in the $J$, $K$, 3.6, 4.5, and $5.7\,\upmu$m bands, where the largest difference consistently occurs at $5.7\,\upmu$m.  An L5 dwarf is the warmest potential companion that can remain undetected in the photometry as a spatially-unresolved source.}
    \label{fig:cool_detect}
\end{figure}

Again using a 10\,per cent photometric excess threshold, various low-mass stellar and substellar objects were added to the measured IRAC fluxes.  Absolute magnitudes for M- and L-type dwarfs were taken from the literature, covering both 2MASS and {\em Spitzer} IRAC wavelengths \citep{Dahn2002,Vrba2004,Patten2006,Leggett2007}, then projected to the appropriate distance, converted to flux and compared to the excess threshold.  It is found that only companions as cool or cooler than an L5 dwarf (approximately 1700\,K; \citealt{Nakajima2004}) would remain undetected (Figure~\ref{fig:cool_detect}).

Third and last, spatially-unresolved white dwarf companions were also considered.  In this case, the contribution of irradiation for the closest orbits such as 2.79\,d is almost certainly irrelevant, as the intrinsic effective temperature of a white dwarf is likely to be significantly higher than the equilibrium temperature of 4970\,K.  The Helix appears to be a thin disk star, with a total space velocity of 35\,km\,s$^{-1}$ based on its radial velocity and proper motion \citep{Durand1998,Gaia2021}, and thus unlikely to be older than a few Gyr.  Absolute magnitudes for white dwarf atmospheric models \citep{Bedard2020} with 0.6\,M$_\odot$ and various temperatures were converted to flux and compared with existing photometry when placed at the appropriate distance.  For temperatures below roughly 30\,000\,K, any spatially-unresolved white dwarf companion would remain undetected; however, the subsequent light curve analysis rules out such an object as the source for the 2.79\,d photometric signal (Section~3).

\subsection{Light curve modeling}

The {\sc lcurve} software package\footnote{\url{https://github.com/trmrsh/cpp-lcurve}} is a sophisticated tool designed to model the light curves of binary systems, especially those including at least one white dwarf.  It achieves this by fitting synthetic light curves to observational data, allowing the derivation of critical parameters of the system \citep{Copperwheat2010}.  For the data considered in this work, the light curve modeling constrains the companion radius and orbital inclination (but not independently).  Phase-folded light curves with 100 evenly-spaced bins are constructed for modelling with {\sc lcurve}, where simulations are run for the 7900\,\AA\ effective wavelength of the {\em TESS} filter.  {\sc lcurve} models both components as blackbodies, an assumption that should minimally affect the results, as any substellar companion would have negligible intrinsic emission, and the hot central star is effectively Rayleigh-Jeans at this wavelength.

It is noteworthy that the light curve modeling is essentially blind to the masses of the two components; however, for white dwarfs the mass is firmly linked to the radius.  Because of this insensitivity, 0.01\,M$_\odot$ is adopted for substellar companions and similarly 0.6\,M$_\odot$ for white dwarf companions.  These simulated masses have no measurable impact on the shape of the modeled light curve, which arises solely from the difference between the day- and night-sides of the companion.  For the {\sc lcurve} simulation inputs, the orbital parameters are derived from Kepler's $3^{\rm rd}$ law, all eclipse-related parameters are disabled, and gravity darkening is set to 0.8 for a convective atmosphere (but is irrelevant for a 2.8\,d orbit).  It is found that a range of limb darkening coefficients for the companion (including limb brightening) have a negligible effect on the results, and thus the value is set to the nearest grid point for linear limb darkening coefficients based on customized calculations for the {\em TESS} bandpass \citep{Claret2017}. Lastly, the absorption factor is set to one, following standard practices for simplifying the model \citep{Parsons2012}.

The {\sc lcurve} analysis took into account the constraints resulting from the lack of photometric excess in the infrared, established in the previous section.  The orbital inclination of the binary system is initialized to match the nebular inclination of $28\degr$ \citep{Henry1999}, and the resulting {\sc lcurve} models are compared against the observed light curve using using the $\upchi^2_\nu$ statistic. 

White dwarf companions are generally far too small to reproduce the observed amplitude of photometric variation via irradiation, regardless of orbital inclination.  While low-mass white dwarfs have relatively large radii that could potentially result in a sufficient light curve amplitude, it would require non-canonical binary evolution to produce a cooler and fainter, yet less massive white dwarf companion.  While a small number of double white dwarfs with age paradoxes are known \citep{Bours2015,Kilic2021}, and where Algol-type evolution likely preceded a common envelope, any white dwarf significantly larger than the Helix central star would yield photometric excess more readily than discussed in Section~2.4.  Thus, while this possibility cannot be ruled out, it is considered unlikely.

For inclinations within this range, {\sc lcurve} modeling determines the corresponding radius that minimizes $\upchi^2_\nu$ when compared to the observations.  Inclination is varied in steps of 1\degr, and the resulting $\upchi^2_\nu$ are weighted by their Gaussian probabilities based on the assumed inclination distribution.  The peak of the resulting distribution indicates the best fit is achieved for a companion radius of 0.021\,R$_\odot$ at $i=25\degr$. The light curve model for this configuration yields $\upchi^2_\nu=1.3$.

\begin{figure}
    \includegraphics[width=\columnwidth]{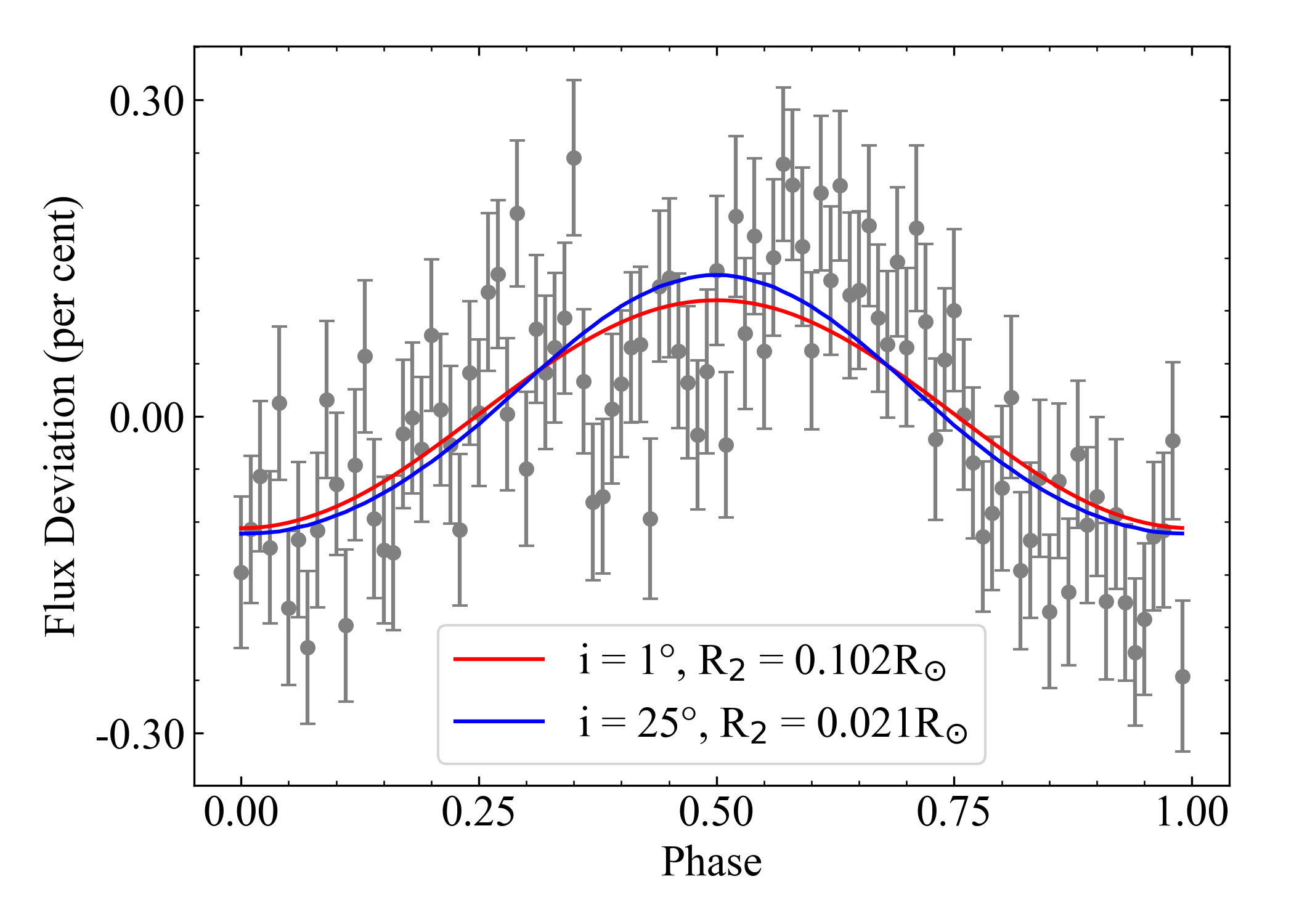}
    \caption{Simulated light curves for the Helix, with two substellar companion models generated by {\sc lcurve}.  The phase-folded light curve is plotted as grey points with error bars, while the models are over-plotted in color.  The  red curve is the best fit resulting from allowing the orbital inclination to vary freely with the companion radius fixed at the maximum allowed based on the SED analysis.  The blue curve results from requiring that the orbital inclination is within $10\degr$ of the nebular inclination, and yields a substantially smaller radius that would correspond to a major exoplanet.  Both models have $\upchi^2_\nu=1.3$.}
    \label{fig:lcurve_fit}
\end{figure}

For the second scenario, the companion radius is fixed at the upper limit of 0.102\,R$_\odot$ as determined from the SED analysis.  For the bulk of possible orbital inclinations, this companion radius results in a light curve amplitude that is too large compared to the data, and thus the optimal solution occurs at the lowest inclination, which is arbitrarily set to $1\degr$ here.  This binary configuration also yields $\upchi^2_\nu=1.3$, and the solutions for both cases are plotted in Figure~\ref{fig:lcurve_fit}.

These two solutions are bench marks within a range of light curve models where the companion radius must monotonically decrease with increasing orbital inclination.  Light curve simulations for binary inclinations higher than that of the nebula are not considered, and would imply even smaller exoplanets; while possible, future confirmation is decreasingly less feasible.

\subsection{Irradiated companion modeling}

Based on the results of the light curve modeling, it is natural to ask if an irradiated, substellar secondary in a 2.79\,d orbit can be detected.  For this goal, the emission spectrum of a potential Jupiter-sized companion is calculated using the {\sc phoenix} self-consistent 1D atmosphere model \citep{Hauschildt1999,Barman2001}, following a setup used for other substellar objects irradiated by white dwarfs \citep{Lothringer2020}, described below. A custom white dwarf atmosphere model for the Helix central star at 120\,000\,K and $\log\,g = 7.47$ is calculated using the Tubingen NLTE Model Atmosphere Package \citep{Rauch2003} to irradiate the companion at a separation of 0.034\,au.

A solar metallicity atmosphere is assumed for the companion, following \citet{Asplund2005}. For convergence, the model is sampled every 1\,\AA\ from 10\,\AA\ to $2.5\,\upmu$m, and at coarser sampling out to $100\,\upmu$m. Higher resolution spectra at $R\approx50\,000$ are then produced from the converged models. Atomic and molecular opacities from all expected major sources are included \citep{Kurucz1995,Hitran2008}. The model is run with 64 vertical layers on a log-spaced optical depth grid from $\uptau=10^{-6}$ to $10^3$. The models were given an internal temperature of 1000~K, though this is negligible compared to the incoming irradiation. Local thermodynamic equilibrium is assumed, but some test models are calculated with hydrogen in NLTE, which exhibit similar emission features.

Two heat redistribution factors are assumed. The first corresponds to planet-wide heat redistribution equivalent to the equilibrium temperature 4970\,K. The second corresponds to dayside-only heat redistribution, which is a factor $2^{1/4}$ hotter than the planet-wide heat redistribution model, coming out to an effective temperature near 6000\,K. Both models exhibit large temperature inversions caused by the absorption of the intense ultraviolet irradiation at high altitudes by the companion atmosphere, similar to those in systems like GD\,245, NN\,Ser, AA\,Dor, and UU\,Sge \citep{Barman2004}, as well as those in substellar companions like WD\,0137$-$349B and EPIC\,212235321B \citep{Lothringer2020} and ultra-hot Jupiters \citep{Fortney2008,Lothringer2018}. These temperature inversions result in an array of atomic emission lines that should be a unique observational fingerprint of an irradiated companion.

\begin{figure}
\centering
    \includegraphics[width=\columnwidth]{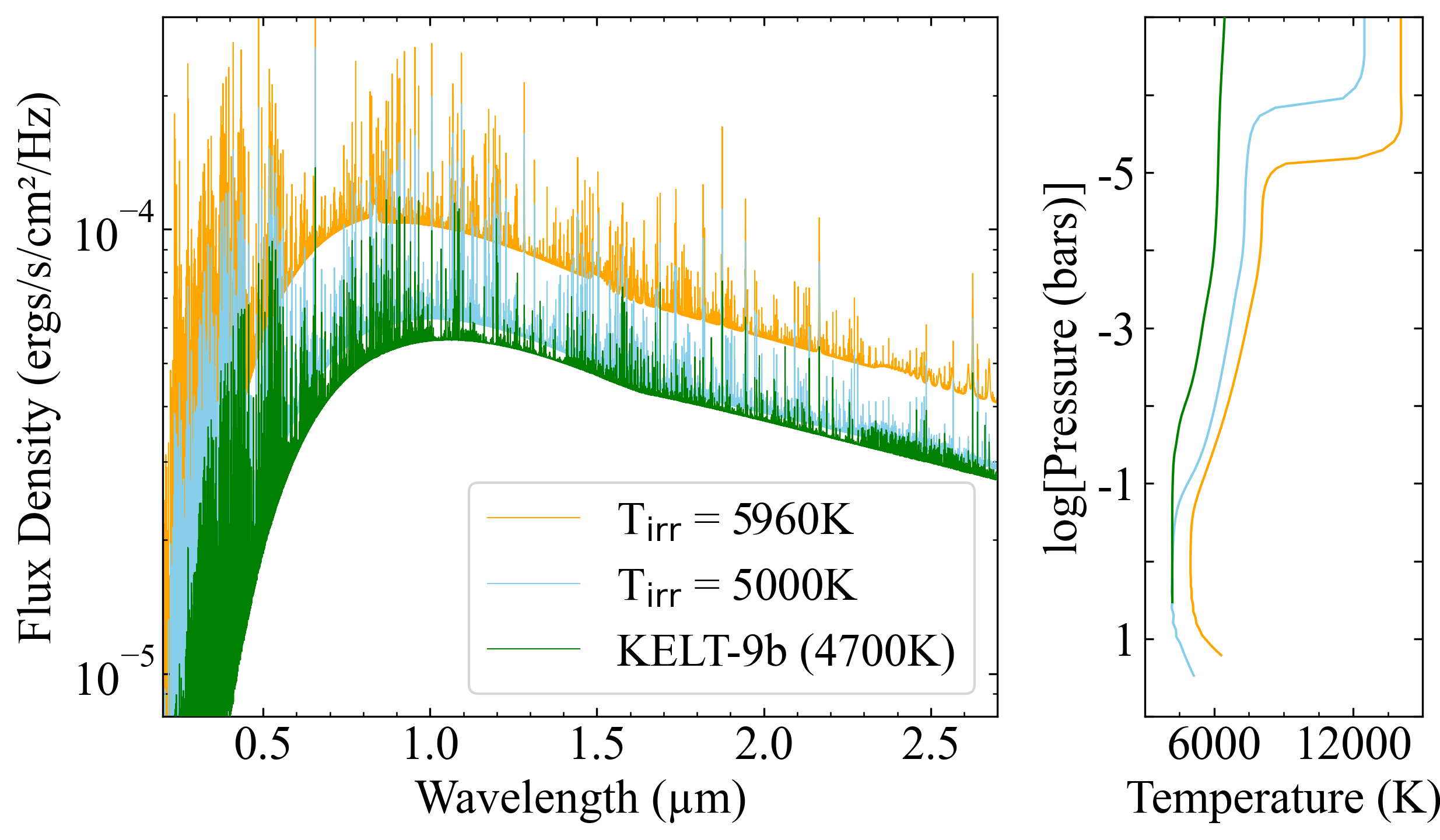}
    \caption{The emergent surface flux for the two irradiated companion models detailed in Section~2.6, and compared to that for KELT\,9b \citep{Lothringer2018}.  While the cooler model with efficient global heat redistribution has an equilibrium temperature comparable to KELT\,9b, the extreme difference in ultraviolet irradiation results in a more strongly-inverted temperature-pressure profile for any planet heated by the Helix.}
    \label{fig:irrmodels}
\end{figure}

\section{Results and Discussion}
\label{sec:3}

The following discusses the implications of the overall constraints, and then focuses on the nature of the 2.79\,d signal that could arise from a companion.

\subsection{Companion constraints within \texorpdfstring{1200\,au}{1200au}}

At orbital distances suitable to influence the nebular morphology, this study suggests that the Helix can only host unseen brown dwarfs or white dwarfs as companions.  Assuming that the progenitor of the Helix central star was a 2\,M$_\odot$ star, a strong binary interaction within a common envelope on the AGB would require the companion to reside within $\sim100-1000$~R$_\odot$ \citep[$\sim0.5-5.0$\,au;][]{Madappatt2016}. Nonetheless, low-mass companions with distant orbits out to roughly 30\,au can play a role in sculpting the mass loss by wind interaction \citep{DeMarco2011}.

In the case of a cooler and fainter white dwarf that would fail to result in photometric excess, canonical evolution predicts such an object would be the more massive of the current pair, and descended from a higher mass main-sequence star than that which evolved into the Helix central star.  The photometric constraints are such that a typical white dwarf need be cooler than 30\,000\,K to remain hidden, but a more massive remnant would be smaller and thus even fainter at the same temperature.  For example, a 1.0\,M$_\odot$ white dwarf at 50\,000\,K would be roughly $20\times$ fainter than the central star and would not exceed the photometric excess threshold, but could have cooled to nearly 12\,000\,K within a Gyr \citep{Bedard2020}.  This is perhaps the strongest possibility if the Helix is currently a binary system.

Substellar companions are more challenging, as they a priori rarely occur both around main-sequence stars \citep{Grether2006,Unger2023}, as well as their white dwarf progeny, where they are straightforward to detect in the infrared for a wide range of substellar masses and virtually all possible orbital separations \citep{Farihi2005,Debes2011}.  That being said, there are notable exceptions and perhaps the Helix is among these; in this case any companion must be a brown dwarf of spectral type L5 or later, where for favorable separations it might be straightforward to directly detect in {\em JWST} imaging observations if it were to lie beyond roughly 0.5\,arcsec = 100\,au in projected separation \citep{DeMarco2022}.  However, it is likely that such orbits are too distant for an architect of the nebular morphology \citep{Soker1997,DeMarco2011}.

There are perhaps two other possibilities for the dark architect of the Helix nebula.  A neutron star might remain hidden in the optical and near-infrared, but would likely be X-ray bright.  For nebular densities of 1000\,cm$^{-3}$ and a flow velocity on the order of 10\,km\,s$^{-1}$, the Bondi-Hoyle accretion rate would lead to X-rays that are over an order of magnitude brighter than those already detected as unresolved from the central star \citep{Guerrero2001}.  A final possibility is a low-mass companion that did not survive the preceding common envelope phase, and merged with the core of the star \citep{soker1998b}.

The {\em Gaia} reduced unit weight error value of 0.95 is not indicative of any duplicity for the Helix central star.  This likely precludes most stellar, and some substellar companions, as well as white dwarfs, to a distance of several au \citep[to periods of roughly 1000\,d;][]{ElBadry2024}.  Thus direct imaging with {\em JWST} should be the next major constraint on hidden companions in wider orbits.

\subsection{Candidate companions with a \texorpdfstring{2.79\,d}{2.79d} orbital period}

\begin{figure}
    \includegraphics[width=\columnwidth]{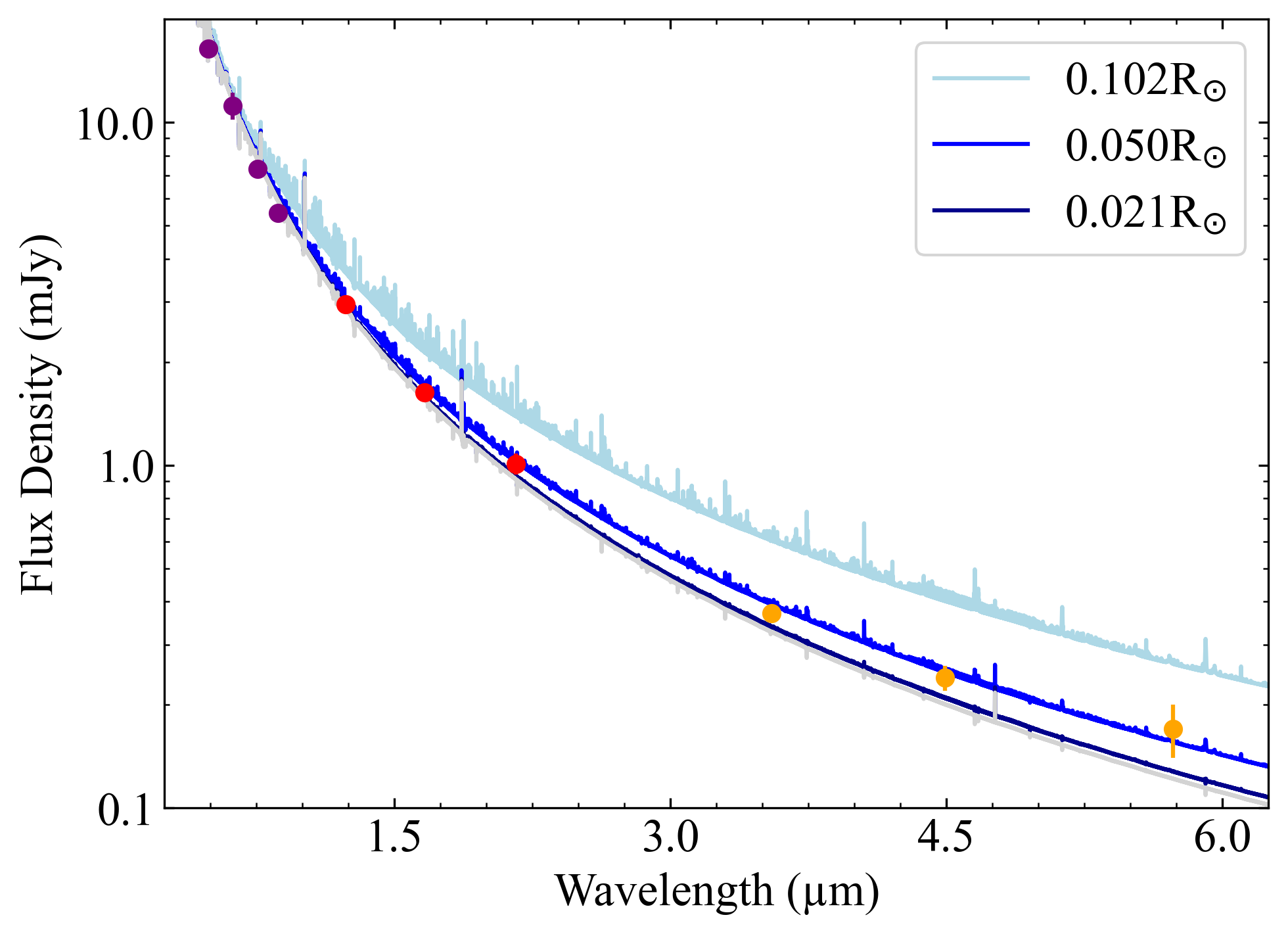}
    \caption{A comparison of the available infrared photometry for the Helix central star with predictions for the combined light of the white dwarf and irradiated companions.  Each plotted model has 4970\,K corresponding to the equilibrium temperature (planet-wide heat redistribution), but the radii vary from 0.102\,R$_\odot$ (light blue), to 0.050\,R$_\odot$ (medium blue), to 0.021\,R$_\odot$ (dark blue).  The irradiated companion flux from a Jupiter-sized body is readily detectable, but for 0.050\,R$_\odot$ or smaller there would be no marked infrared excess (see Section~3.2). It should be noted that the $5.7\,\upmu$m IRAC photometry may exhibit a slight excess from the cool dust more strongly detected at longer wavelengths \citep{Su2007}.}
    \label{fig:wdplusirr}
\end{figure}

The primary goal of this study is to ascertain the properties of any companion with a 2.79\,d orbital period that may be responsible for the observed light curve variability detected with {\em TESS} \citep{Aller2020}.  While the lack of photometric excess in the SED does not exclude the possibility of a white dwarf companion in such an orbit, the {\sc lcurve} analysis rules this out.  Combining these two analyses yields an upper radius limit for of 0.102\,R$_\odot$ for an unevolved companion.

The irradiated atmosphere modeling provides further constraints on the size of any companion by comparing the predicted flux at infrared wavelengths with existing photometry.  In Figure~\ref{fig:wdplusirr} are plotted three companion models that yield a range of predicted fluxes above and below the current photometry.  For this exercise, the fainter model with the planet-wide heat redistribution was selected, as it is fainter and thus more conservative in terms of radius constraints.  As can be seen in the plot, a Jupiter-sized body with equilibrium temperature 4970\,K would result in photometric excess not only in {\em Spizer} IRAC photometry but also at 2MASS wavelengths, and can be ruled out.  Given that this solution requires an arbitrarily low orbital inclination to satisfy the light curve amplitude constraints, this independent constraint is reassuring.

The light curve modeling solution with orbital inclination $i=25\degr$ and companion radius 0.021\,R$_\odot$ is likely to be the most realistic.  First, it is based on the observed tendency for orbits to be aligned with nebular inclinations for well-studied binaries in planetary nebulae \citep{Hillwig2016}.  Second, such an exoplanetary companion would remain hidden, where the irradiated companion model re-scaled to this radius is consistent with existing photometry.  However, such an exoplanet could be somewhat larger and still remain undetected; Figure~\ref{fig:wdplusirr} shows that a companion re-scaled to 0.050\,R$_\odot$ would also remain consistent with infrared photometry.  Such a radius would require a lower orbital inclination to remain consistent with the observed light curve, but is allowed.

\begin{figure}
\centering
    \includegraphics[width=\columnwidth]{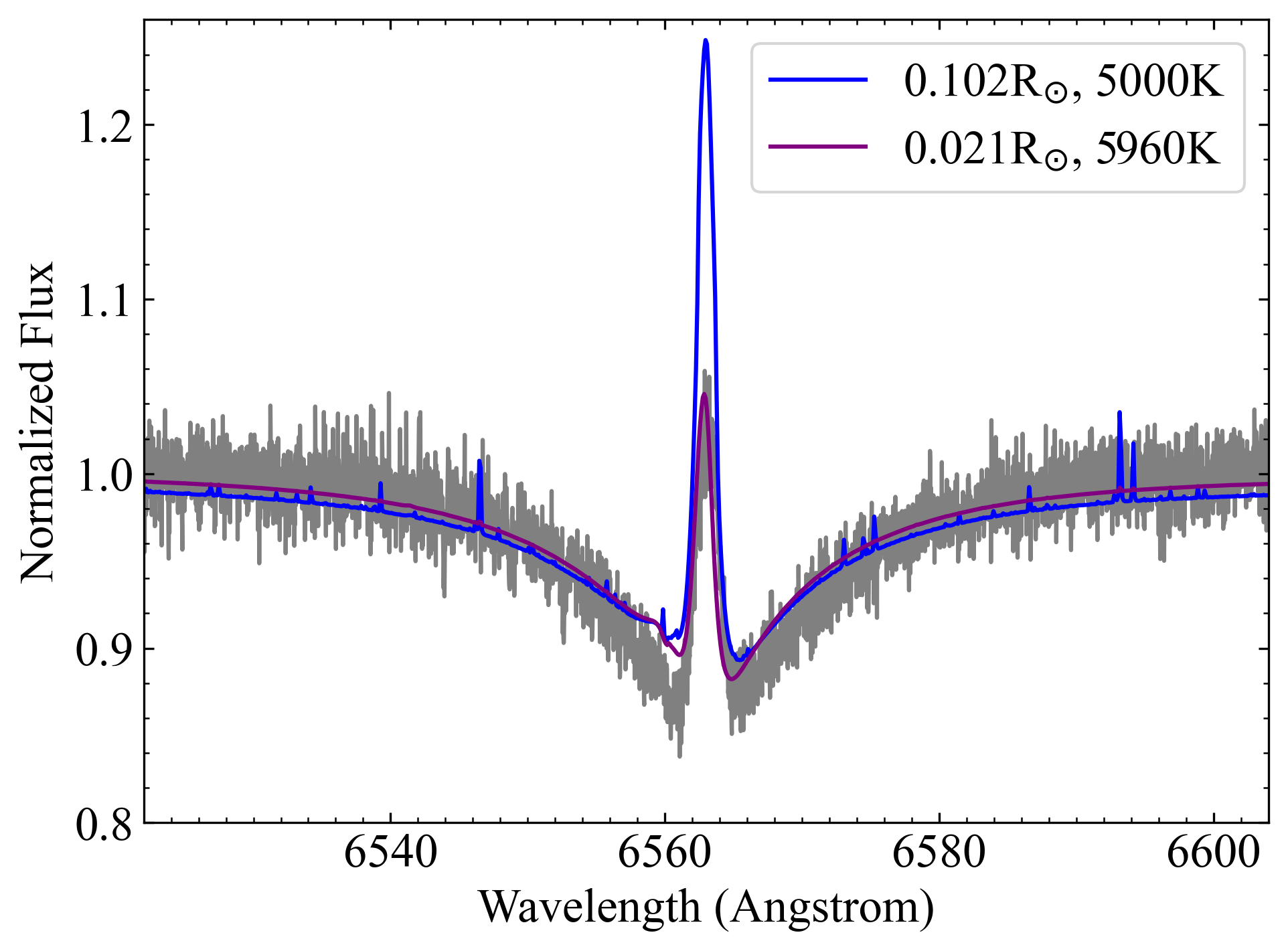}
    \caption{Comparison of spectral data and models in the region of H$\upalpha$ for the Helix central star.  Plotted in grey is a spectrum taken with the VLT and UVES instrument on 2002 Oct 12, with signal-to-noise around 75 and resolving power $R\approx50\,000$.  The data exhibit a strong emission feature in the core of the H$\upalpha$ absorption feature which is caused by NLTE effects in the upper atmosphere.  Overplotted are the combined model fluxes of the white dwarf photosphere and one of two irradiated atmospheres: a 0.102\,R$_\odot$ body contributes more flux than observed even at the lower $T_{\rm irr}$, while a 0.021\,R$_\odot$ planet does not contribute any appreciable flux, even at the higher $T_{\rm irr}$.}
    \label{fig:uves}
\end{figure}

Another, modest constraint on the companion can be estimated from the strength of the observed H$\upalpha$ emission line in the optical spectrum of the Helix central star, where an irradiated companion may contribute emission in addition to the intrinsic, NLTE stellar atmosphere feature.  While there are numerous emission lines predicted based on the irradiated atmosphere modeling, the H$\upalpha$ line will almost certainly be the strongest in the optical range, relative to the white dwarf atmosphere (Figure~\ref{fig:irrmodels}).  For comparison with the models, there are several spectra of WD\,2226$-$210 in the European Southern Observatory (ESO) archive, where the observation with the highest signal-to-noise is selected and plotted in Figure~\ref{fig:uves} (Program 70.C-0100, PI Kendall). 

While there does not appear to be any evidence for additional emission, there are two considerations for the comparison of these models and observations, both related to the phase of the companion orbit.  First, a Jupiter-mass companion in a 2.79\,d orbit would have a circular velocity of 133\,km\,s$^{-1}$, and could cause a radial velocity shift of up to 3\,\AA\ between its emission component and the stellar emission, thus enabling a more straightforward detection at some orbital phases.  Second, and less favorable to detection is the fact that emission lines naturally appear and disappear with the corresponding day and night sides of the illuminated companions; this is evident in numerous white dwarf plus M dwarf binary systems, where some extreme examples are known \citep{Maxted1998,Parsons2010,Hallakoun2023}.

There are six UVES spectra of WD\,2226$-$210 covering H$\upalpha$ in the ESO archive, all taken between 2001 and 2008, where none display any visual evidence of additional emission beyond that predicted by the model atmosphere.  However, it may be better to obtain spectral data in the region of Pa$\upalpha$ or longer wavelength hydrogen transitions, where the irradiated atmosphere is predicted to have relatively strong infrared emission, but where the white dwarf is considerably fainter than in the optical.  Infrared wavelengths are  dominated by telluric emission and absorption and would likely require space-based observations; there are currently {\em JWST} observations of the Helix central star using MIRI MRS, but not at shorter wavelengths where emission lines are most prominent based on Figures~\ref{fig:irrmodels} and \ref{fig:wdplusirr}.

\subsection{Additional considerations for the light curve signal}

While a tantalizing possibility, the identification of a candidate exoplanet within the Helix Nebula requires further evidence if it is to be considered viable.  An exoplanet in an extreme environment and subject to such intense irradiation is a radical departure from the solar system and the bulk of known exoplanets.  In particular, it is not expected for a planet to survive the dense stellar envelope when the host star is on the AGB, where the nominal expectation is destruction by ablation and tidal disruption after sufficient orbital contraction \citep{Nordhaus2010,Staff2016,Lau2022,Oconnor2023}.

Beyond the shape of the nebula itself, the only current evidence that suggests a companion of any kind is the light curve variation every 2.79\,d, where it is the low amplitude of this photometric variability that supports a small companion radius.  The shape of the phase-folded light curve deviates somewhat from purely sinusoidal behavior, which is possible for strongly irradiated companions in close orbits (e.g.\ owing to tidal distortion or Doppler beaming), but not expected to play a role for a 2.79\,d orbit.  It is possible the non-sinusoidal deviations in the phased light curve are simply the result of a weak signal that has been observed for only a handful of cycles, and that may be affected by {\em TESS} de-trending routines because of the relatively long period.

Another possibility is some type of star spot, where the photometric period would correspond to the rotation rate.  While not currently understood, there are a number of high temperature white dwarfs with variable light curves, although these are stars with either ultra-high energy excitation lines or overly strong He\,{\sc ii} lines \citep{Reindl2021}, neither of which apply to the Helix central star.  These photometrically variable, hot white dwarfs are hypothesized to be magnetic, and thus form a small class compared to the field, and none are currently associated with planetary nebulae.  The light curve of the Helix does not share the same morphology as that observed for the ultra-high excitation stars, whose light curves are typically characterized by a single peak and a wide and low trough (see figs~3-5 in \citealt{Reindl2021}).  Lastly, the period of 2.8\,d lies at the extreme end of the resulting distribution for those light curves, where the median is around 0.6\,d.

However, none of these factors rule out a stellar origin for the light curve variability observed toward the Helix central star.  One possibility would be an inhomogeneous distribution of metals on the stellar surface, similar to that observed in chemically peculiar stars \citep[e.g.][]{Krtivka2015,Prvak2015}.


\section{Conclusion}
\label{sec:4}

This study has placed new constraints on potential companions to the Helix Nebula, and specifically any that would have a sufficiently close orbit to be the unseen architect of the nebular structures.  The prime focus of the study is the candidate companion responsible for the 2.8\,d periodic signal in the {\em TESS} light curve \citep{Aller2020}, with a supplementary aim to place constraints on spatially-unresolved companions that would otherwise be detected via photometric excess. 

The results of the light curve modeling rule out all but substellar companions as possible sources of any irradiation-modulated brightness variations.  If the inclination is ignored, then a Jupiter-sized (0.102\,R$_\odot$) brown dwarf provides a good fit to the light curve, but requires a close to a face-on inclination.  However, an irradiated atmosphere model for such a companion would yield an infrared excess that is not observed.  If the inclination of the companion orbit is similar to the nebular inclination, then a significantly smaller body is necessary to fit the low-amplitude light curve variation, resulting in a radius of 0.021\,R$_\odot$, and which would correspond to an exoplanet.  Such a small companion would remain undetected in existing photometry and spectroscopy.  However, some type of star spot cannot be ruled out at present, and if applicable to the Helix, then other light curves of planetary nebulae central stars might also be the result of stellar surface inhomogeneities.

Prospects for confirming or ruling out an exoplanetary companion are challenging at best.  There are no dimming events in the {\em TESS} light curves, but even a grazing transit would be readily detected for a size ratio $R_{\rm p}/R_\star\approx0.8$.  Radial velocity monitoring is also problematic: the central star has no narrow line cores in either absorption or emission, and with $M_1/M_2 = K_2/K_1 \approx 0.001$ for a Jupiter-mass companion, the white dwarf would have a velocity semi-amplitude no greater than 0.1\,km\,s$^{-1}$ (cf.\ $\upDelta v\approx2$\,km\,s$^{-1}$ for ESPRESSO; \citealt{Pepe2014}).  Near-infrared $1-5\,\upmu$m spectroscopy with {\em JWST} might prove more sensitive to emission lines from an irradiated companion in a wavelength region where the central star contributes significantly less than in the optical.

Should it remain plausible that a low-mass brown dwarf or exoplanet orbits the Helix central star in such close proximity, it raises questions about the survival of planetary systems post-main sequence.  There are a number of dynamical and hydrodynamical studies for this type of evolution, and all predict direct engulfment and destruction via ablation or tidal disruption \citep{Mustill2012,Nordhaus2013,Staff2016,Lau2022}.  There is at least one likely exoplanet in a close orbit around a white dwarf, but the 1.4\,d orbit of this system is likely to have resulted from Lidov-Kozai migration after the post-main sequence evolution of the host star, as it is part of a hierarchical stellar triple \citep[WD\,1856+534;][]{Vanderburg2020,Munoz2020}.  Otherwise, the survival of an exoplanet in such a close orbit may require fine tuning; to avoid destruction, the object would need to enter the AGB envelope only within the latter-most portion of the final thermal pulse \citep{Oconnor2023}.  That being said, some studies have shown that engulfment of at least one massive exoplanet can contribute to asymmetries observed towards the Helix and similar nebulae \citep{Keaveney2020}.

If instead the light curve signal is stellar in nature, then this study demonstrates that only a fainter, more massive, and cooler white dwarf can persist at the orbital separations necessary to shape the Helix.  It is possible such a companion might be too distant to be detected via {\em Gaia} astrometry, and in this case high-resolution imaging with {\em JWST} or future ground-based facilities may probe this possibility further.  In this case, the architect of the Helix nebula remains hidden, or was destroyed in the act of creation.

\section*{Acknowledgements}

This paper includes data collected by the {\em TESS} mission, which are publicly available from the Mikulski Archive for Space Telescopes.  This work is based (in part) on observations made with the {\em Spitzer Space Telescope}, which was operated by the Jet Propulsion Laboratory, California Institute of Technology under a contract with NASA.

\section*{Data Availability}
The data used in this study are available in public archives.


\bibliographystyle{mnras}
\bibliography{ref} 


\bsp    

\end{document}